\input{aipcheck}

\documentclass[
    ,final            
  ]
  {aipproc}

\layoutstyle{8x11single}

\pdfoutput=1

\begin{document}

\title{Higgs searches at Tevatron}

\classification{}
\keywords      {Higgs, Tevatron}

\author{Lars Sonnenschein}{
  address={LPNHE, Univerit\'es Paris VI et VII}
}



\begin{abstract}


SM and MSSM Higgs Searches at the proton anti-proton collider
Tevatron in Run~II are presented. The performance of the collider and 
the two experiments D0 and CDF is shown. No deviation from SM background 
expectation and no MSSM Higgs signal has been observed.  
\end{abstract}

\maketitle


\section{Introduction}


The Higgs boson is the last missing particle in the Standard Model. Direct searches at
LEP2 set a lower limit of $m_H>114.4$~GeV at 95\% confidence level (C.L.) on its mass.
Further constraints by mass measurements of the top quark and 
the $W$ boson and electro-weak parameter fits of the LEP Electro Weak Working Group 
suggest a rather light Standard Model Higgs boson.
The electro-weak fit yields $m_H<175$~GeV at 95\% C.L., which extends to
$m_H<207~$GeV if the LEP2 limit is included. 

Higgs production cross sections at the proton antiproton collider Tevatron 
are small. They reach $0.1-1$~pb depending on the Higgs mass.
For Higgs masses below 135~GeV the Higgs is predominantly decaying into
a $b\bar{b}$ quark pair.
The dominant production mechanism $gg\rightarrow H$ suffers here an overwhelming
mulit-jet background.
Searches can be performed with lower background in the associated Higgs production   
including a $W$ or $Z$ boson in the final state.
The most promising channels are $WH\rightarrow\ell\nu b\bar{b}$ and 
$ZH\rightarrow\nu\bar{\nu}b\bar{b}$.
For Higgs masses above 135~GeV the predominant decay channel is $H\rightarrow WW^{(*)}$
where at least one $W$ boson is on-shell.
Preferentially the leptonic decay channels $gg\rightarrow H\rightarrow WW^{(*)}\rightarrow \ell^+\nu\ell^-\bar{\nu}$ and $WH\rightarrow WWW^{(*)}\rightarrow\ell^{\pm}\ell^{\pm}+X$ can be explored.

\section{Experimental environment}

\subsection{Tevatron collider}
In comparison to Run~I, in which proton antiproton collisions have been taken place
at the Tevatron collider at a center of mass energy of $\sqrt{s}=1.8\,\mbox{TeV}$,
in Run~II the center of mass energy has been elevated
to $\sqrt{s}=1.96\,\mbox{TeV}$, giving rise to 10~\% higher Higgs production cross sections.
The bunch spacing has been reduced from $3500\,\mbox{ns}$ to
$396\,\mbox{ns}$.
Major upgrades to the Linac and main injector 
together with a new antiproton recycler and electron cooling made it possible
to achieve in Run~IIa instantaneous luminosities of up to 
$170 \cdot 10^{30}\mbox{cm}^{-2}\mbox{s}^{-1}$.                                                                                                    
\subsection{Detector upgrades}
The CDF and D$\emptyset$ detectors have been massively upgraded for Run~IIa. 
Driven by physics goals they became very similar. Beside a replaced 
silicon tracker and central drift chamber the
geometric acceptance of the CDF detector has been increased by new forward
calorimeters and extended muon coverage. A new silicon tracker, new preshower detectors
and a $2\,\mbox{T}$ superconducting solenoid have been added to the
D$\emptyset$ detector. Its muon coverage has been extended as well.
The data acquisition and trigger systems have been upgraded for both detectors
to cope with the shorter bunch spacing compared to Run~I.

Of major importance for the Run~IIb upgrade 
is the layer zero (L0) microvertex detector
of D\O\ which is by the time of this writing inserted and fully read out. Its 
most remarkable feature is the signal to noise performance of $S/N=18$.

\subsection{Data samples}
The Run~II physics data taking started in 2002 (February for CDF and July for
D$\emptyset$). The reported Higgs search results are
based on data taken until November 2005 and vary depending on the analysis
channel between 261 and $950\,\mbox{pb}^{-1}$. This has to be compared to about
$120\,\mbox{pb}^{-1}$ accumulated in Run~I.
                                                                                                    
\section{Standard Model Higgs searches}

\subsection{$Z/\gamma^*\rightarrow ee$ + jets (D\O)}
The understanding and rejection of background is crucial in Higgs searches.
Therefore the associated production of jets along with a pair of charge conjugated electrons
originating form a $Z/\gamma^*$ is investigated.
The event selection encompasses two electrons above a transverse momentum of 25~GeV 
of which one has to be central ($|\eta|<1.1$),
any number of jets with a transverse energy above 15~GeV and an invariant di-electron mass of
$70<m_{e^+e^-}<120$~GeV.
The jet-multiplicity distribution of data is compared to simulation.
While the event generator PYTHIA 6.319 shows more events in higher jet-multiplicity bins
starting form three jets on, the matrix elements form the program Sherpa, which are matched
to a parton shower via the CKKM mechanism, show good agreement for jet-multiplicity bins
$N_{\mbox{jet}} \leq 4$.

\subsection{$ZH\rightarrow \nu\bar{\nu}b\bar{b}$ (CDF)}
In the $ZH$ production considered here a Higgs is radiated from an off-shell $Z$ boson which
then becomes on-shell and decays into a neutrino antineutrino pair while the
Higgs decays into a $b\bar{b}$ pair of heavy quarks.
The crucial elements of the event selection consist of 
two jets with a transverse momentum above 60 and 25~GeV, of which at least one has to be 
$b$-tagged. To account for the two neutrinos which leave the collision undetected a
missing transverse energy above 70~GeV is requiered. Backgrounds are $W/Z$ production 
associated with heavy flavour jets, multi-jet production, di-boson production,
mistagged $b$-jets and $t\bar{t}$ pair production. For a hypothesized Higgs mass of 120~GeV
the events inside an invariant jet-jet-mass window between 80 and 120~GeV are counted.  
289~pb$^{-1}$ of intergated luminosity are used.
Six events are observed and $4.36\pm1.02$ events are predicted. This leads to a 95\% C.L.
exclusion limit of 4.5~pb.

\subsection{$ZH\rightarrow \nu\bar{\nu}b\bar{b}$ (D\O)}
The $ZH$ analysis of D\O\ has an improved event selection with respect to the former analysis.
It requires two acoplanar jets with a transverse energy above 20~GeV, missing transverse energy 
above 50~GeV and the scalar sum of the jets transverse energy to be below 240~GeV.
The analysis is done for single and double $b$-tagged events separately to subdivide
channels of highly different sensitivities. After derivation of the limits both channels
are combined taking correlated uncertainties into account.
For a hypothesized Higgs mass of 115~GeV the events inside an invariant jet-jet-mass window
between 75 and 125~GeV are counted. An integrated luminosity of 261~pb$^{-1}$ has been used.
11 events are observed and $9.4\pm1.8$ are predicted, leading to an exclusion limit of
4.3~pb at 95\% C.L.

Using this analysis, limits on $WH$ production with a missed charged lepton can be placed.
This improves the combined limits on $WH$ production.

\subsection{$WH\rightarrow \ell\nu b\bar{b}$ (CDF)}
The $WH$ analysis considered here expects either an electron or a muon and a neutrino
from an on-shell $W$ boson and a $b\bar{b}$ quark pair from the Higgs.
To select events of interest the jets have to have a transvers momentum above 15~GeV,
an electron or a muon above 20~GeV transverse momentum and missing transverse energy above
20~GeV. The analysis is done separately for single and double $b$-tagged events. 
The sensitivity of single $b$-tagged events has been increased by exploiting
a Neural Network (NN) $b$-tagger which uses the information of different $b$-tagging 
variables, which are typically not 100\% correlated. The double $b$-tagged events are chosen
with a secondary vertex tagger since no improvement in sensitivity could be obtained in using 
the NN $b$-tagger. The analysis is based on an integrated luminosity of 695~pb$^{-1}$.
332 events with at least one $b$-tagged jet are observed. 
The background consists of mistagged events, $Wb\bar{b}$, $Wc\bar{c}$, $Wc$, $t\bar{t}$,
single top, di-boson and multi-jet production. The prediction of total background amounts to 
$318.8\pm54.7$ events. Limits are derived by fitting the di-jet invariant mass spectrum.
For a hypothesized Higgs mass of 115~GeV an upper exclusion limit of 3.6~pb at 95\% C.L.
has been established.

\subsection{$WH\rightarrow \ell\nu b\bar{b}$ (D\O)}
Here, the $WH$ analyses are done separately in the electron and the muon channel
(plus the missed lepton channel as mentioned above in the $ZH$ analysis). Further
the single and double $b$-tagged events are treated separately
and then combined later to drive a unique limit based on all decay channels and 
$b$-tag sub-samples. While the muon channel has been analysed for the first time at D\O\
the electron channel has been re-optimized. The event selection is based on one central 
isolated electron or muon with a transverse energy above 20~GeV. The missing transverse 
energy has to exceed 25~GeV and exactly two jets above a transverse energy of 20~GeV and
within a pseudorapidity of $|\eta|<2.5$ have to be found. At least one of them has to be 
$b$-tagged. The limit is derived from counting the number of events inside an invariant
jet-jet-mass window around a hypotesized Higgs mass of 115~GeV. The integrated luminosity
used varies slightly for the different $WH$ search channels and ranges from 371 to 
385~pb$^{-1}$. 32 single $b$-tagged events and six double $b$-tagged events are observed. 
The number of predicted events amounts to $45.1\pm6.9$ and $9.3\pm1.8$. 
A combined exclusion limit of 2.5~pb at 95\% C.L. can be established.

\subsection{$t\bar{t}H\rightarrow \ell +2q +4b$ (CDF)}
The Higgs production associated with a top antitop quark pair is investigated
in the $t\bar{t}$ lepton plus jets channel where the top quarks decay into a $b$ quark
and a $W$ boson, of which one decays leptonically and the other one hadronically, giving rise
to a light $q\bar{q}$ pair. The Higgs is decaying into a $b\bar{b}$ quark pair.
So there are altogether six jets expected in an event, amoung which four are supposed 
to be $b$-tagged. To distinguish the signal from background - which consists dominantly of
mistagged events, irreducible processes with the same final state and multi-jet background 
- exactly one identified electron or muon is required. At least five jets with a transverse 
momentum above 15~GeV, amoung which at least three jets have to be $b$-tagged, are required.
Missing transverse energy has to exceed 25~GeV to account for the neutrino. 
An integrated luminosity of 320~pb$^{-1}$ has been used for this analysis.
One event is being observed while $0.89\pm0.12$ background events are predicted.
The exclusion limit on the associated production process for a hypothesized Higgs mass of 115~GeV 
is 0.95~pb at 95\% C.L.

\subsection{$WH\rightarrow WWW^*\rightarrow \ell^{\pm}\ell^{\pm} +X$ (D\O)}
The $WH$ production with three $W$ bosons (of which at least two are on-shell) is
considered in the final state with two like sign leptons, i.e. one lepton from the Higgs
and one lepton form the associated $W$ boson. In this way the background of $Z$ bosons decaying 
into a pair of charge conjugated same flavour leptons can be removed very efficienctly.
The two like sign leptons are required to be isolated and having a transverse momentum above
15~GeV. In addition, missing transverse energy above 20~GeV is demanded to account for neutrinos.
An integrated luminosity of 636~pb$^{-1}$ has been used.
In the $ee$-channel one event is observed while $0.70\pm0.08$ are predicted.
In the $e\mu$-channel three events  are observed versus $4.32\pm0.23$ events predicted
and finally in the $\mu\mu$-channel two events are observed while $3.72\pm0.75$ are predicted.
For a hypothesized Higgs mass of 115~GeV an exclusion limit of 3.88~pb could be set.

\section{$H\rightarrow WW^*\rightarrow \ell^+\ell^-\nu\bar{\nu}, \ell=e,\mu$ (CDF)}

The Higgs invariant mass cannot be reconstructed in the case of leptonical 
$W$ boson decay since the neutrinos leave the collision undetected. On the other hand
one can exploit the spin correlation between the two leptons to discriminate background
where two lepton candidates are not originating via a $W$ boson from a scalar particle
like the SM Higgs boson. The spins of the two $W$ bosons originating from a scalar
Higgs boson have to add up to zero. Due to the left-handed character of the electro-weak
interaction ($W^{+(-)}$ bosons are only left (right) circular or longitudinal polarized)
the two leptons tend to be emitted collinear. To select the events two charge conjugated 
leptons are required. One of them has to have a transverse 
momentum above 20~GeV and the other above 10~GeV. Missing transverse energy has to exceed
one forth of the hypothesized Higgs mass for which a limit is being derived.
The invariant dilepton mass has to exceed 16~GeV and to be smaller than $m_H-5$~GeV.
An integrated luminosity of 360~pb$^{-1}$ has been used.
The difference in azimuthal angle between the two leptons is histogrammed and fitted to derive a limit.
For a hypothesized Higgs mass of 120~GeV the exclusion limit at 95\% C.L. is 4.5~pb.

\subsection{$H\rightarrow WW^*\rightarrow \ell^+\ell^-\nu\bar{\nu}, \ell=e,\mu$ (D\O)}
The D\O\ analysis requires in the same final state two charge conjugated isolated leptons 
with a transverse momentum above 20 and 15~GeV. A veto on the $Z$ resonance and energetic 
jets is applied. An integrated luminosity of 950~pb$^{-1}$ has been used.
For a hypothesized Higgs mass of 120~GeV 31 events are observed and 
$32.7\pm2.3$ are predicted, leading to an exclusion limit of 6.3~pb at 95\% C.L.

\section{Standard Model Higgs limits}
\begin{figure}[h]
  \hspace*{-8ex}
  \includegraphics[height=0.5\textheight]{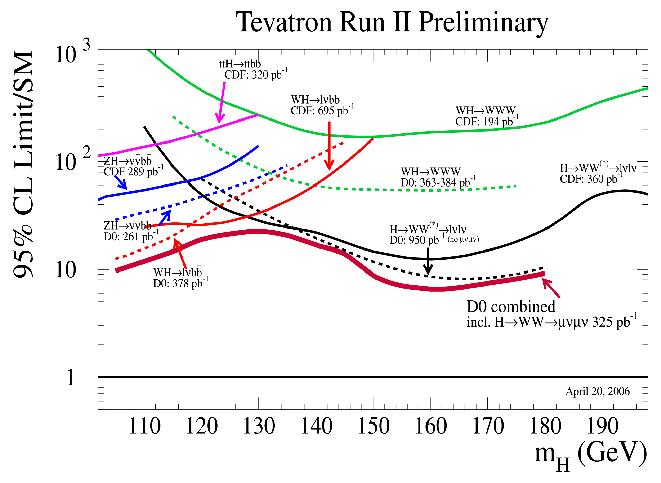}
  \caption{\label{smlimits} D\O\ and CDF Higgs exclusion limits normalized to the cross section
           predicted by the SM}
\end{figure}

Exclusion limits normalized to the SM cross section of all discussed analyses versus 
Higgs mass are shown in fig. \ref{smlimits}. 
In the case of D\O\  a combination of all search channels is given, too.
A combination of all Tevatron results, including those of CDF, is in progress.

\vspace*{2ex}

\pagebreak

\section{Minimal Supersymmetric SM Higgs}

\subsection{Supersymmetric Particles and parameters}

Five scalar particles are predicted in the MSSM: $h$, $H$, $A$, $H^+$, $H^-$. In CP-conserving models
$h$ and $H$ are CP-even while $A$ is CP-odd. At tree-level perturbation theory there are two
independent parameters: the ratio of the vacuum expectation values $\tan\beta$ and the mass
$m_A$. Five more parameters intervene via radiative corrections as shown in table  
\ref{susyparameters}. In supersymmetry one could have a light Higgs but with small couplings.
$2\times 2$ benchmark scenarios are studied \cite{Carena2005}, which are maximal and no-mixing scenarios 
for a positive and negative Higgs mass parameter $\mu$.
\begin{table}[h]
\begin{tabular}{c|l|l|l}
 Parameter & Description & $m_h$-max & no-mixing \\ \hline
 $M_{\mbox{\small SUSY}}$ & Parameterizes squark, gaugino masses & 1~TeV & 2~TeV \\
 $X_t$ & Related to trilinear coupling $A_t$ ($\rightarrow$ stop mixing) & 2~TeV & 0 \\
 $M_2$ & Gaugino mass term & 200~GeV & 200~GeV \\
 $\mu$ & Higgs mass parameter & $\pm200$~GeV & $\pm200$~GeV \\
 $m_{\mbox{\small gluino}}$ & enters via loops & 800~GeV & 1600~GeV 
\end{tabular}
\caption{\label{susyparameters} Susy parameters through radiative corrections and benchmark scenarios.}
\end{table}

\subsection{Neutral MSSM Higgs $\rightarrow \tau\tau$ (D\O\  and CDF)}
\begin{figure}[h]
  \hspace*{26ex}
  \includegraphics[height=.59\textheight]{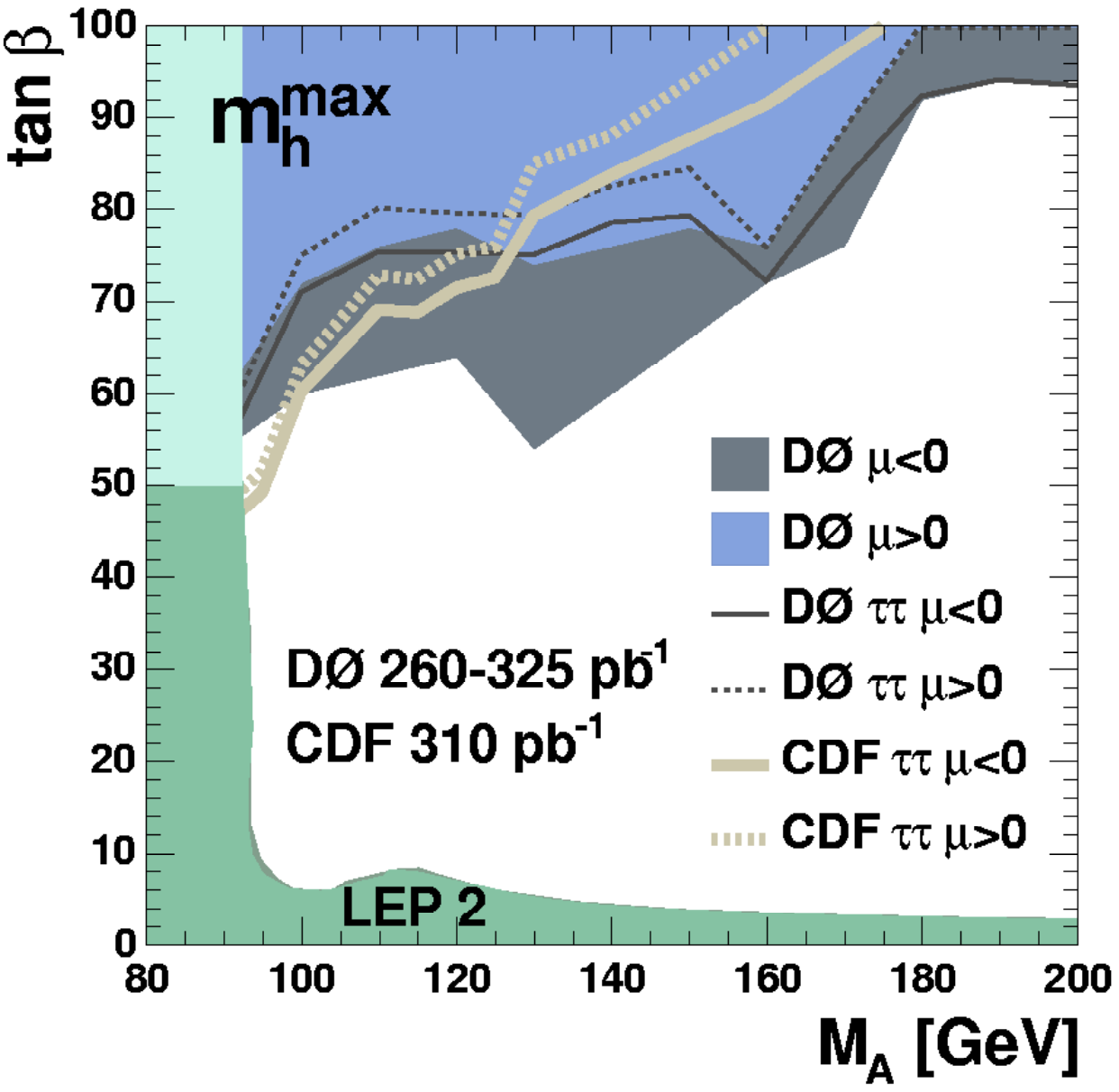}
  \hspace*{-20ex}
  \includegraphics[height=.59\textheight]{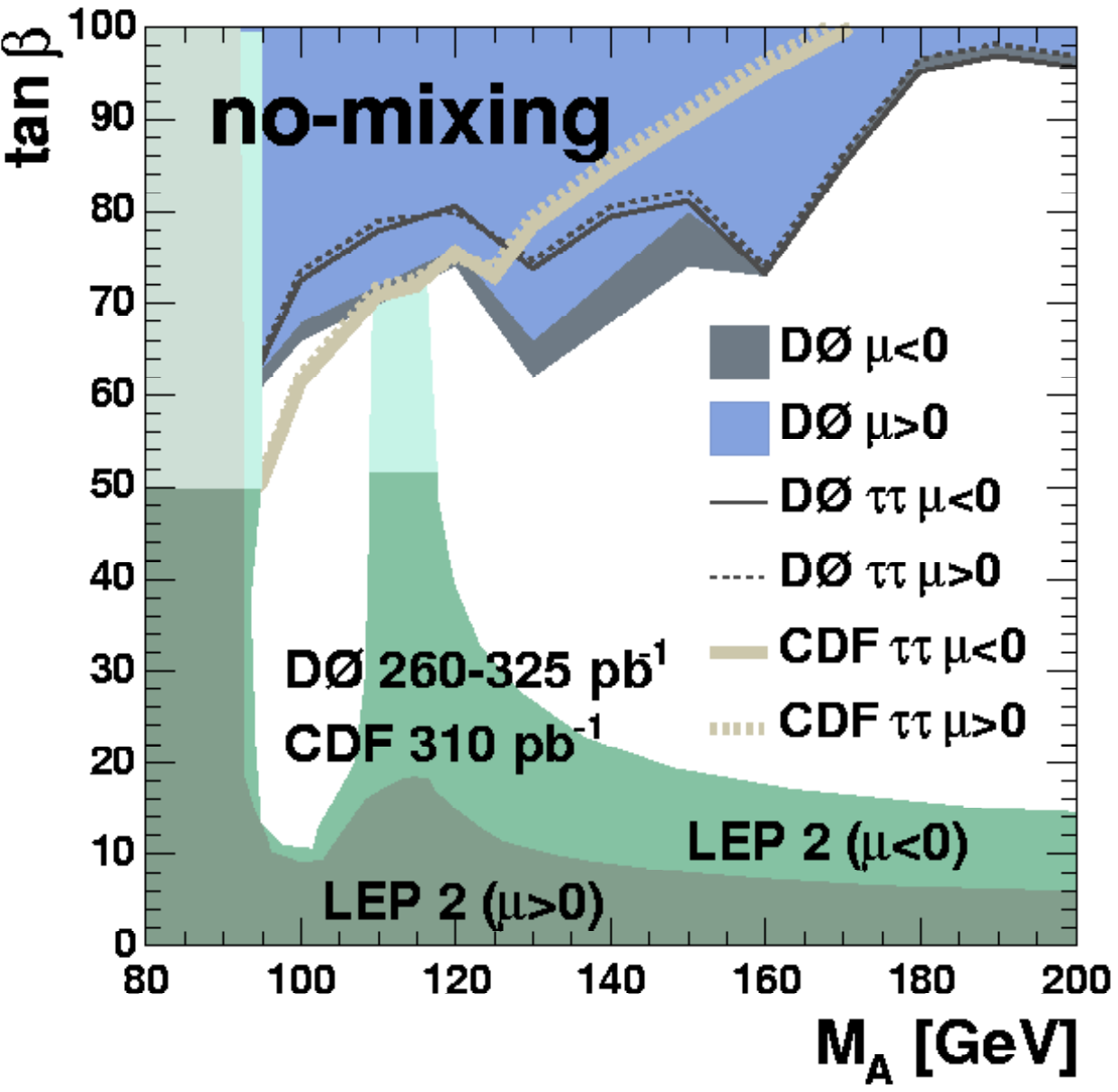}
\vspace*{-9ex}
 \caption{\label{mssmlimits} D\O\ and CDF Higgs into $\tau\tau$ exclusion limits 
 in the $\tan\beta$ versus $m_A$ plane for maximal mixing (left) and no-mixing (right) 
scenarios. Limits are indicated for both signs of the Higgs mass parameter $\mu$. The results
contain all $\tau$ decay channels.}
\end{figure}

The signature of signal events is given by two reconstructed tau leptons and missing 
transverse energy. In the case of D\O\ different hadronic tau decay topologies are
distinguished from each other and from quark and gluon induced jets 
by means of a neural network. Standard model backgrounds are $Z\rightarrow\tau\tau$
which is irreducible, $Z/\gamma^*\rightarrow ee,\mu\mu$, multi-jet production,
$W\rightarrow\ell\nu$ and di-boson production.
\begin{table}[h]
\begin{tabular}{c|c|c|c}
Experiment & Channel & Data & Expected background \\ \hline
     & $e+\tau_{\mbox{\small had}}$ & 484 & $427.3\pm55.3$ (stat$\oplus$sys$\oplus$lum) \\
D\O\ & $\mu+\tau_{\mbox{\small had}}$ & 575 & $576.3\pm61.5$ (stat$\oplus$sys$\oplus$lum) \\
     & $e+\mu$ & 42 & $43.5\pm5.3$ (stat$\oplus$sys$\oplus$lum) \\ \hline
CDF  & $\tau_{1,2} \rightarrow X$ & 487 & $496\pm54$(stat)$\pm27.7$(sys)$\pm24.8$(lum) \\
\end{tabular}
\caption{\label{mssmeventnumbers} 
Observed and expected event yields in Higgs to $\tau\tau$ search.}
\end{table}
D\O\ used an integrated luminosity of 325~pb$^{-1}$ and CDF used 310~pb$^{-1}$.
To derive limits the visible mass 
$M_{\mbox{\small vis}}=\sqrt{P_{\mbox{\small vis}}(\tau_1)+P_{\mbox{\small vis}}(\tau_2)+E\!\!\!\!\!/\;_T}$ is being fitted.
The numbers of observed and expected background events are given in table \ref{mssmeventnumbers}.
Fig. \ref{mssmlimits} shows the exclusion plots in the plane of $\tan\beta$ and $m_A$ for
maximal mixing and no-mixing scenarios. LEP 2 limits are indicated as well.
\vspace*{-45ex}

\subsection{Charged MSSM Higgs (CDF)}
\begin{figure}[h]
  \includegraphics[height=.54\textheight]{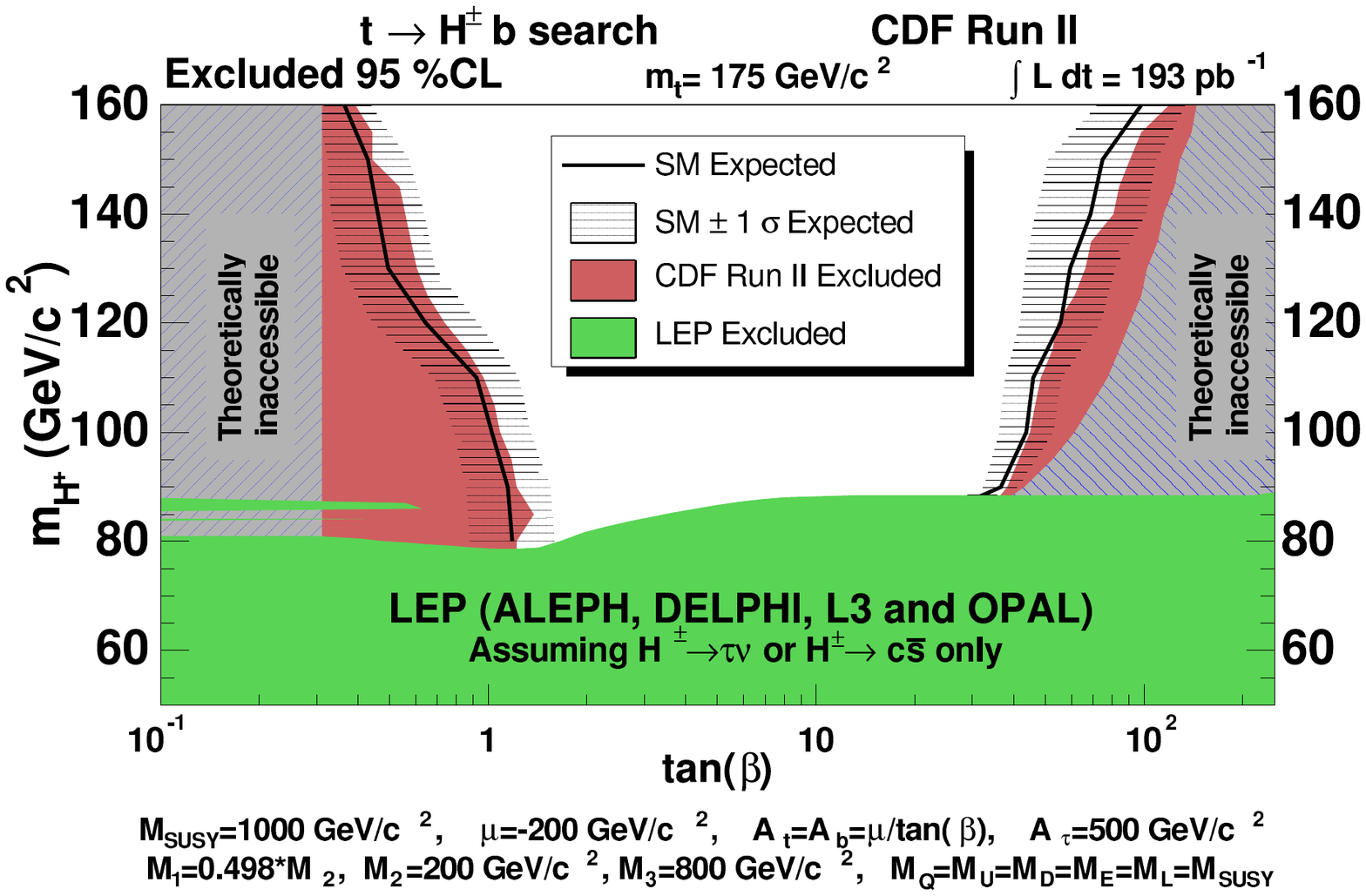}
  \hspace*{-2ex}
  \includegraphics[height=.54\textheight]{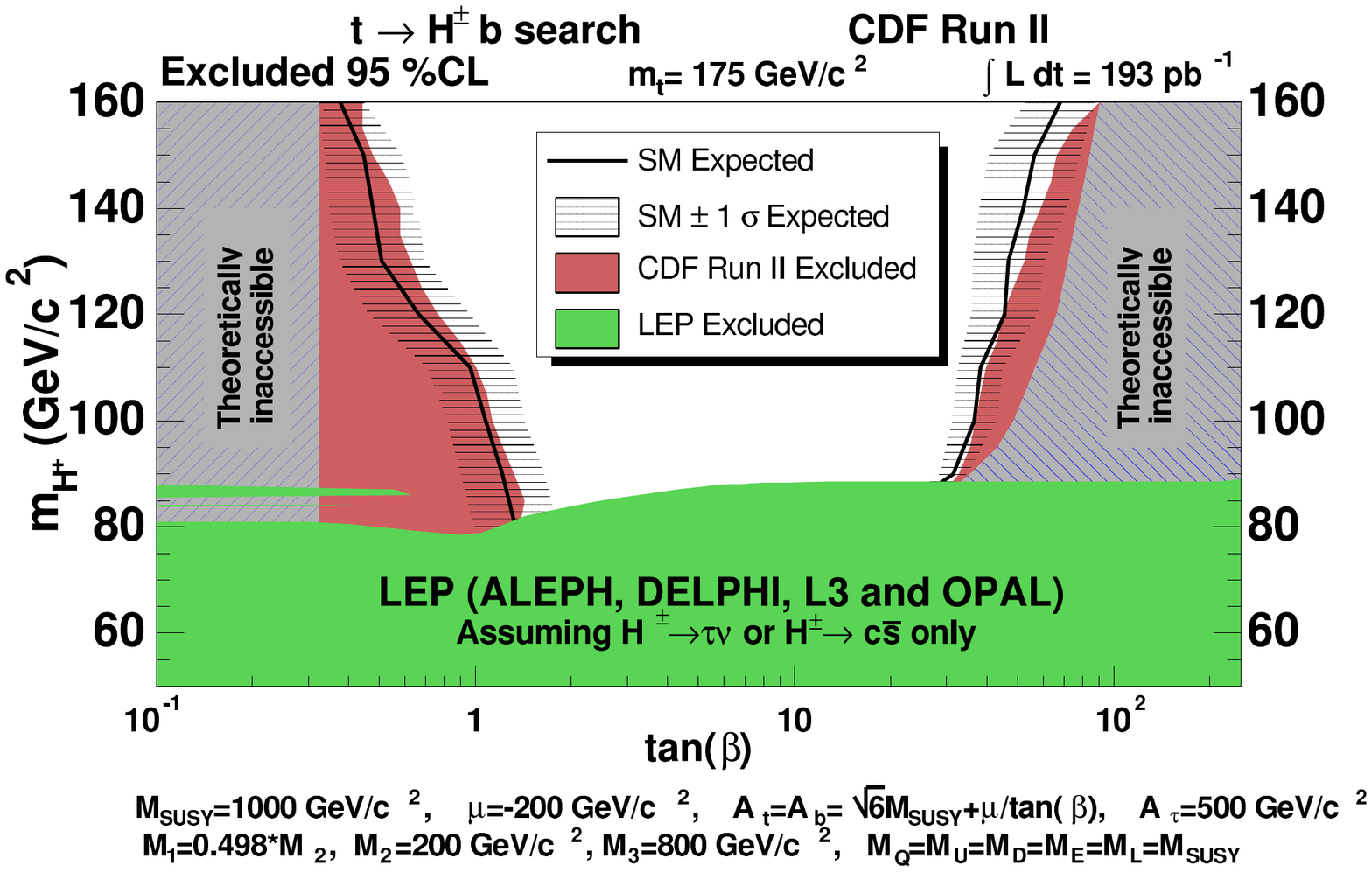}
\put(-378,75){\bfseries\sf No mixing}
\put(-150,75){\bfseries\sf $m_h$-max mixing}
\vspace*{-14ex}
 \caption{\label{mssmchargedlimits} CDF exclusion limits for a charged Higgs originating from 
 a top quark, assumed to decay exclusively into $\tau\nu$ or $cs$. The $m_{H^+}$ versus 
$\tan\beta$ plane is shown for no-mixing (left) and maximal mixing (right) scenarios.}
\end{figure}

The SM top quark decays predominantly into a $W$ boson and a $b$ quark.
The $W$ boson of the top quark decay could be replaced by a charged Higgs boson
if its mass is smaller than the top quark mass (subtracted by the $b$ quark mass).
The $t\bar{t}$ production is exploited in search for a charged Higgs, whose
major decay modes are $H\rightarrow \tau\nu$, $cs$, $t^*b(\rightarrow)Wbb$ and 
$Wh (\rightarrow Wb\bar{b})$. The branching ratios depend on $\tan\beta$ and $m_{H^+}$ 
and they are different from $W$ boson branching ratios. An integrated luminosity
of 193~pb$^{-1}$ has been used. Final states with leptons and jets are selected as
indicated by the different decay channels in table \ref{chargedhiggseventnumbers},
where also the numbers of observed and expected events are given. The result is consistent
with the SM. In fig. \ref{mssmchargedlimits} the exclusion limits for a charged Higgs 
originating from a top quark, assumed to decay exclusively into $\tau\nu$ or $cs$ are
shown in the $m_{H^+}$ versus $\tan\beta$ plane. The left plot shows the no-mixing scenario
while the right plot shows the maximal mixing scenario. Exclusion limits from LEP experiments are 
also indicated.

\begin{table}[h]
\begin{tabular}{l|c|c|c}
Channel & Data & Expected background & SM expectation ($t\rightarrow Wb$) \\ \hline
 $2\ell+$jets           & 13 &  $2.7\pm0.7$ & 11 \\ 
 $\ell+$jets ($1b$)     & 49 & $20.3\pm2.5$ & 54 \\
 $\ell+$jets ($\geq2b$) & 8 &  $0.94\pm0.1$ & 10 \\
 $\ell+\tau+$jets       & 2 &   $1.3\pm0.2$ &  2
\end{tabular}
\caption{\label{chargedhiggseventnumbers} Observed and expected event yields in charged Higgs 
into $\tau\nu$, $cs$ search. Non $t\rightarrow Wb$ expected background is indicated in a separate
column.}
\end{table}
\vspace*{-42ex}

\section{Conclusions}
Tevatron and experiments are performing well. A wide range of Higgs searches has been performed
by both, CDF and D\O\ experiments up to an integrated luminosity of 1~fb$^{-1}$ in Run~II.
No deviation from the SM background expectation has been observed and no signal has been observed 
in MSSM Higgs search. Work to improve sensitivity is under way. A first combination of all SM
Higgs channels has been presented by D\O. Combination efforts between D\O\ and CDF have started.


\begin{theacknowledgments}
Many thanks to the stuff members at Fermilab, collaborating institutions
and the Higgs physics group convenors of the CDF and D$\emptyset$
experiments for their support and help. 
\end{theacknowledgments}



\bibliographystyle{aipproc}   


\begin{thebibliography}{9}








\bibitem{Carena2005}
M. Carena {\it et al.}, hep-ph/051123, 2005


\end{thebibliography}




\end{document}